\documentclass[conference]{IEEEtran}

\usepackage{amsmath}
\usepackage{amsfonts}
\usepackage{amssymb}
\usepackage{algorithm}
\usepackage{algorithmic}
\usepackage{graphicx}
\usepackage{verbatim}
\usepackage{multicol}

\newtheorem{theorem}{Theorem}

\newtheorem{example}{Example}

\newcommand{\beq}{\begin{equation}}
\newcommand{\eeq}{\end{equation}}
\newcommand{\beqnn}{\begin{equation*}}
\newcommand{\eeqnn}{\end{equation*}}
\newcommand{\beqy}{\begin{eqnarray}}
\newcommand{\eeqy}{\end{eqnarray}}
\newcommand{\beqynn}{\begin{eqnarray*}}
\newcommand{\eeqynn}{\end{eqnarray*}}
\newcommand{\bit}{\begin{itemize}}
\newcommand{\eit}{\end{itemize}}
\newcommand{\ben}{\begin{enumerate}}
\newcommand{\een}{\end{enumerate}}
\newcommand{\bex}{\begin{example}}
\newcommand{\eex}{\end{example}}

\newcommand{\balg}[1]{\begin{algorithm} \caption{#1}}
\newcommand{\ealg}{\end{algorithm}}

\newcommand{\balgc}{\begin{algorithmic}[1]}
\newcommand{\ealgc}{\end{algorithmic}}

\newcommand{\bary}{\begin{array}}
\newcommand{\eary}{\end{array}}
\newcommand{\bmx}{\begin{bmatrix}}
\newcommand{\emx}{\end{bmatrix}}
\newcommand{\bsmx}{\left[\begin{smallmatrix}}
\newcommand{\esmx}{\end{smallmatrix}\right]}
\newcommand{\bmxc}[1]{\left[\begin{array}{@{}#1@{}}}
\newcommand{\emxc}{\end{array}\right]}
\newcommand{\bcn}{\begin{center}}
\newcommand{\ecn}{\end{center}}

\renewcommand{\v}{\boldsymbol{v}}

\usepackage[rgb]{xcolor}

\usepackage{slashbox}

\usepackage{psfrag}
\usepackage{amssymb}
\usepackage{amsmath}
\usepackage{cite}
\usepackage{graphics}
\usepackage{graphicx}
\usepackage{epsfig}
\usepackage{subfigure}
\usepackage{url}
\usepackage{amscd}
\usepackage{threeparttable}
\usepackage{enumerate}
\usepackage{multirow}

\begin{document}

\title{SMDP-based Downlink Packet Scheduling Scheme for Solar Energy Assisted Heterogeneous Networks}

\author{\IEEEauthorblockN{
Qizhen~Li\IEEEauthorrefmark{1}\IEEEauthorrefmark{2},
Jie~Gao\IEEEauthorrefmark{2},
Jinming~Wen\IEEEauthorrefmark{3},
Xiaohu~Tang\IEEEauthorrefmark{1},
Lian~Zhao\IEEEauthorrefmark{2}, and
Limin~Sun\IEEEauthorrefmark{4}}\\
\vspace{-3mm}
\IEEEauthorblockA{\small \IEEEauthorrefmark{1}School of Information Science and Technology, Southwest Jiaotong University,\\ Chengdu 611756, China (e-mail: xhutang@swjtu.edu.cn)}
\IEEEauthorblockA{\small \IEEEauthorrefmark{2}Department of Electrical and Computer Engineering, Ryerson University,\\ Toronto M5B 2K3, Canada (e-mail: \{qizhen.li, j.gao, l5zhao\}@ryerson.ca)}
\IEEEauthorblockA{\small \IEEEauthorrefmark{3}Department of Electrical and Computer Engineering, University of Toronto,\\ Toronto M5S 3G4, Canada (e-mail: jinming.wen@utoronto.ca)}
\IEEEauthorblockA{\small \IEEEauthorrefmark{4}Beijing Key Laboratory of IoT Information Security Technology, Institute of Information Engineering,\\ Chinese Academy of Sciences, Beijing 100093, China (e-mail: sunlimin@iie.ac.cn)}
}

\maketitle

\begin{abstract}
Renewable energy assisted heterogeneous networks can improve system capacity and reduce conventional energy consumption. In this paper, we propose a semi-Markov decision process (SMDP)-based downlink packet scheduling scheme for solar energy assisted heterogeneous networks (HetNets), where solar radiation is modeled as a continuous-time Markov chain (CTMC) and the arrivals of multi-class downlink packets are modeled as Poisson processes. The proposed downlink packet scheduling scheme can be compatible with the mainstream wireless packet networks such as long-term evolution (LTE) networks and the fifth-generation (5G) networks because the SMDP is a real-time admission control model. To obtain an asymptotically optimal downlink packet scheduling policy, we solve the semi-Markov decision problem using the relative value iteration algorithm under average criterion and the value iteration algorithm under discounted criterion, respectively. The simulation results show that the average cost of the SMDP-based packet scheduling scheme is less than that of the greedy packet scheduling scheme.
\end{abstract}

\section{Introduction}
The explosive growth of mobile data traffic has led to the significant increase of energy consumption in wireless communication networks. In recent years, energy harvesting communications have attracted great attention from both academic and industrial research communities, because energy harvesting technologies can shift power supply from fossil fuels to renewable energy sources, e.g., solar radiation, wind, tides, etc.\cite{ku2016advances}. However, one of the main challenges in energy harvesting communications is the renewable energy scheduling and allocation due to the randomness of renewable energy and mobile data traffic.

Heterogeneous networks (HetNets) with mixed macro-cell and small cell deployment are the main architectures of the fifth generation (5G) mobile networks to improve system capacity \cite{wang2014cellular}. HetNets assisted by energy harvesting technologies can reduce the conventional energy (such as grid power) consumption \cite{zhang2017energy, dhillon2014fundmentals}. A few researchers have proposed some renewable energy scheduling and allocation methods in energy harvesting HetNets. Some of the works modeled energy harvesting HetNets as slot-based systems \cite{chiang2016green, han2016provisioning, wei2018user}. In \cite{chiang2016green} and \cite{wei2018user}, the authors assumed that the base stations have infinite data to be sent and proposed strategies to maximize energy efficiency. In \cite{han2016provisioning}, the authors averaged the transmission capacities of base stations with the given arrival rates and the amount bits of packets in some locations. In \cite{zhang2016energy} and \cite{han2016energy}, the authors proposed the energy harvesting HetNets based on real-time admission control, where the variation of battery is modeled as an M/D/1 queue and the traffic intensity is modeled as a Poisson point process (PPP) in \cite{zhang2016energy}, and both the energy arrivals and the downlink packet arrivals are modeled as Poisson counting processes in \cite{han2016energy}.

The radio resource scheduling period of the existing communication protocols is on the order of milliseconds, e.g., the resource scheduling period of LTE networks is $1ms$,  while intensity change period of green energy is at least on the order of seconds. Therefore, it is difficult to apply slot-based joint optimization algorithms to practical communication networks. The existing real time methods also have difficulty in finding optimal energy scheduling policies. A semi-Markov decision process (SMDP)\cite{puterman1994markov} and a continuous-time Markov decision process (CTMDP) \cite{guo2009continous} are efficient ways to model the resource management problem in real-time admission control energy harvesting systems \cite{murtaza2013optimal, martinez2016maximum, liu2015optimal}. An SMDP-based optimal data transmission and battery charging policy for solar powered sensor networks was proposed in \cite{murtaza2013optimal}. In \cite{martinez2016maximum}, a routing algorithm based on the SDMP was proposed in wireless sensor networks, where the arrivals of energy harvest and traffic packets are modeled as Poisson processes. A CTMDP-based optimal threshold policy for in-home smart grid with renewable generation integration was proposed in \cite{liu2015optimal}.

Generally, energy harvesting is modeled as a point process in existing literatures. In this paper, solar energy assisted HetNets are considered, where a macro base station (MBS) is powered by the power grid and a small base station (SBS) is powered by solar radiation. We model solar radiation as a continuous-time Markov chain (CTMC) and model the arrivals of multi-class downlink data traffic as Poisson processes. The downlink packet transmission link selection problem is modeled as an SMDP, which can be compatible with the mainstream wireless packet networks. We solve the semi-Markov decision problem using the relative value iteration algorithm under average criterion and the value iteration algorithm under discounted criterion, respectively.

\section{System Model}
\subsection{System Description}
\begin{figure}[H]
  \centering
  \includegraphics[width=7.0cm,height=4.2cm]{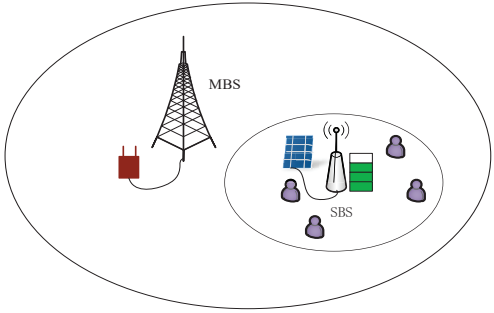}\\
  \caption{The system model of solar energy assisted HetNets}\label{system model}
\end{figure}

Solar energy assisted HetNets considered in this paper are shown in Fig. \ref{system model}. We assume that communication resources (e.g., frequency bandwidth, spacial channel and non-orthogonal multiple access (NOMA) channel) are abundant, i.e., every packet can be allocated enough communication resources in time. Users in a small cell can connect to an SBS and an MBS, where the SBS is powered by solar radiation and the MBS is powered by the power grid. The SBS connects the MBS via high-speed wired link.

\subsection{Energy Model}
We model the intensity of solar radiation on the ground based on cloud cover. The number of solar radiation states is assumed to be $R+1$. The intensity of solar radiation at state $r$ is $G_r$ (in $watt/m^2$), where $0 \leq r \leq R$. The radiation state sojourn time is determined analytically by the cloud size and wind speed $v_w$ (in $m/s$). The wind speed is assumed to be stationary for a long time. We assume that the cloud size is exponentially distributed with mean $d_r$ (in $m$, i.e., the mean diameter of the cloud that results in solar radiation state $r$). Therefore, the evolutions of solar radiation states can be regarded as a CTMC where the state sojourn time follows exponential distribution. For example, the transitions among the solar radiation states are sequential and circular in \cite{niyato2007sleep}. The transition rate matrix corresponding to the CTMC can be expressed as follows:
\begin{equation}\label{}
  \textbf{R}_c=\left(
  \begin{array}{cccc}
    -\frac{v_w}{d_0} & \frac{v_w}{d_0} &  &  \\
     & -\frac{v_w}{d_1} & \frac{v_w}{d_1} &  \\
     & \ddots & \ddots &  \\
    \frac{v_w}{d_R} &  &  & -\frac{v_w}{d_R} \\
  \end{array}
\right),
\end{equation}
where the expected sojourn time of state $r$ is $\tau_r=d_r/v_w$, correspondingly, the transition rate is $\beta_r=1/\tau_r$.

We assume that the area of the photovoltaic panel on the SBS is $\Omega_S$ (in $m^2$) and the conversion efficiency of solar energy is $\eta$. Given solar radiation state $r$, the charging power is:
\begin{equation}\label{}
  p_r=\eta G_r \Omega_S.
\end{equation}

We assume the battery in the SBS has limited capacity $E$ (in $Joule$). The minimum unit of energy that the battery can process (charge and discharge) is $E_{min}$ (in $Joule$). The maximum amount of energy units in the battery is:
\begin{equation}\label{}
  M=\left\lfloor \frac{E}{E_{min}} \right\rfloor.
\end{equation}
where $\lfloor x \rfloor$ represents rounding $x$ to the nearest integer towards minus infinity. Therefore, the battery state is defined as $m$ ($0\leq m \leq M$).

\subsection{Traffic Model}
We only consider the downlink data traffic. It is reasonable for the solar power assisted HetNets because transmitting signals needs to consume much more energy than receiving signals. One of the most prominent features of the now available and future wireless networks is supporting multi-service applications, such as voice, video, web browsing, file transmission, interactive gaming, etc.\cite{yang2007optimizing}. Therefore, the downlink packets sent to the users in the small cell are classified into $N$ classes. The arrival process of class $n$ packet is assumed to be a Poisson process with arrival rate $\lambda_n$, where $1 \leq n \leq N$. The arrival processes of all the classes of packets are assumed to be independent. For a class $n$ packet, it takes $\zeta_n$ units of energy to transmit a packet from the MBS to a user and takes $\xi_n$ units of energy to transmit a packet from the SBS to a user. In general, $\zeta_n > \xi_n$.

\section{SMDP-based Packet Scheduling Model}
The downlink packet scheduling process is modeled as an SMDP in which the distribution of the time to the next decision epoch and the state at that time depend on the past only through the state and action chosen at the current decision epoch \cite{puterman1994markov}. In our SMDP model, the time intervals between adjacent decision epochs follow exponential distributions. Generally, an SMDP can be formulated as a 5-tuple, i.e., $\{t_i, \mathcal{S}, \mathcal{A}, Q, c\}$, where $t_i$, $i\in \mathbb{N}$ (symbol $\mathbb{N}$ denotes the set of non-negative integers) is an decision epoch, $\mathcal{S}$ is the state space, $\mathcal{A}$ is the action space, $Q$ is the transition probability which includes the state transition probability and the state sojourn time distribution, and $c$ is the immediate cost. In the rest of this section, we will formulate the SMDP in detail for the considered problem.

\subsection{Decision Epoch}
In the SMDP, the time interval between two adjacent decision epochs can be a duration with a random length within $[0, \infty)$, so that downlink packets can be sent timely.
\subsection{State Space}
In the SMDP model, a decision-making state is considered, which includes a conventional state $\hat{s}$ and an event $e$, i.e., $s=<\hat{s}, e>$. The conventional state includes the solar radiation state $r$ and the battery state $m$, denoted as $\hat{s}=[r,m]$. The arrival event of class $n$ packet can be defined as $e_n$, $1 \leq n \leq N$. The next solar radiation state is defined as $r'$. Thus, the arrival event of the next solar radiation state is defined as $e_{r'}$. The state space is the set of all the available decision-making states, represented as:
\begin{equation}\label{}
  \mathcal{S}=\{s|s=<\hat{s},e>\},
\end{equation}
where $e \in \{e_1,\ldots,e_N,e_{r'}\}$.

\subsection{Action Space}
The downlink packets can be transmitted by the MBS or SBS (if it has enough energy in the battery). The action that a packet is transmitted by the MBS is defined as $a_s=0$ and the action that a packet is transmitted by the SBS is defined as $a_s=1$. When the solar radiation state changes, the controller takes an fictitious action, defined as $a_s=-1$. The action space is defined as the set of all possible actions, as follows:
\begin{equation}\label{}
  \mathcal{A}=\{a_s|1,0,-1\}.
\end{equation}

\subsection{State Transition Probability}
Given solar radiation state $r$, the amount of time to harvest a unit energy is:
\begin{equation}\label{}
  T_r=\frac{E_{min}}{p_r}.
\end{equation}
The harvested energy is accumulated after the decision epoch. Assuming that the current battery state satisfies $m<M$ and the action is $a=0$ (or $a=-1$), the next battery state $m'$ should be one of the states in the states set $\{m,m+1,\ldots,M\}$. Correspondingly, the next event occurs in the time intervals $\{[0,T_r),[T_r,2T_r),\ldots,[(M-m)T_r,\infty)\}$ from the current decision epoch.

\begin{figure}[H]
  \centering
   \includegraphics[width=7.0cm,height=6cm]{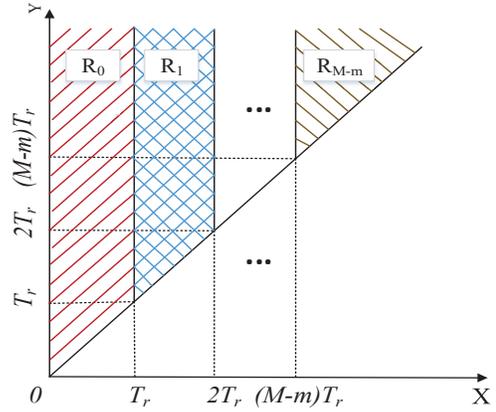}\\
  \caption{State transition probability}\label{transition}
\end{figure}

If the next event is the arrival of a class $n'$ packet (i.e., $e_{n'}$) and occurs within duration $[0,T_r)$, the next state is $s'=<[r,m],e_{n'}>$. Since all the downlink packet arrivals and the next solar radiation state arrival are independent under the given state-action pair $(s,a)$, the occurrence rate of the next event is:
\begin{equation}\label{}
  \gamma(s,a)=\sum_{n=1}^N \lambda_n+\beta_r,
\end{equation}
where the time interval from the state-action pair $(s,a)$ to the next event $e_{n'}$ occurrence follows an exponential distribution with parameter $\lambda_{n'}$ (represented as the random variable $X$ in Fig. \ref{transition}), and the time interval to the other events occurrence follows an exponential distribution with parameter $\sum_{n\neq n'}\lambda_n+\beta_r$ (represented as the random variable $Y$ in Fig. \ref{transition}). The probability density functions are respectively as follows:
\begin{equation}\label{}
  f_X (x)=\lambda_{n'} e^{-\lambda_{n'} x},
\end{equation}
\begin{equation}\label{}
  f_Y(y)=(\sum_{n\neq n'}\lambda_n +\beta_r)e^{-(\sum_{n\neq n'}\lambda_n +\beta_r)y}.
\end{equation}
Random variables $X$ and $Y$ are independent. Since the next battery state is $m'=m$, the state transition probability (the integral of the function $f_X(x)\cdot f_Y(y)$ over the region $R_0$ in Fig. \ref{transition}) is:
\begin{equation}\label{}
  \begin{split}
  &P_a(s'|s)=\int_0^{T_r}\int_x^{\infty} f_X(x)\cdot f_Y(y) dydx \\
  &~~~~~~~~=\int_0^{T_r}f_X(x)\left[\int_x^{\infty}f_Y(y)dy\right]dx \\
  &~~~~~~~~=\int_0^{T_r} \lambda_{n'}e^{-\gamma x}dx \\
  &~~~~~~~~=\frac{\lambda_{n'}}{\gamma}(1-e^{-\gamma T_r}).
  \end{split}
\end{equation}

In Fig. \ref{transition}, $R_2$ represents the integral area of the probability that $e_{n'}$ precedes the other events in the interval $[T_r, 2T_r)$ and $R_{M-m}$ represents the integral area of the probability that $e_{n'}$ precedes the other events in the interval $[(M-m)T_r, \infty)$.  Using the same method, all the state transition probabilities can be obtained.

\subsection{Immediate Cost}
Since the solar power equipment needs a certain cost and the charging and discharging processes also have some loss to equipment, we assume the price of every unit of solar energy is $\omega_s$. We also assume the price of every unit of grid power energy is $\omega_m$. Usually, $\omega_s < \omega_m$. We consider the energy costs consumed at decision epochs as immediate costs:
\begin{equation}\label{}
  \begin{split}
c(s,a)=\begin{cases}
            \omega_m\zeta_n,~~~a_s=0 \\
            \omega_s\xi_n,~~~~a_s=1 \\
            0,~~~~~~\mathrm{others}.
          \end{cases}
\end{split}
\end{equation}

\section{Packet Scheduling Policy Optimization}
The simplest link selection policy is the greedy policy, i.e., when a packet arrives, it is served by the SBS as long as there exists enough energy in the battery; otherwise the packet is served by the MBS. However, the greedy policy may not minimize the long-term energy costs. For example, there are two class packets. If $\omega_m\zeta_1-\omega_s\xi_1 > \omega_m\zeta_2-\omega_s\xi_2$ and $\xi_1<\xi_2$, the action $a_s=0$ may be better than $a_s=1$ when event $e_2$ occurs.  Therefore, we should utilize the statistical characteristics to get a long-term optimal packet scheduling policy. In this section, we present two semi-Markov decision problem formulation criterions to get asymptotically optimal link selection policies; one is the average criterion, and the other is the discounted criterion.
\subsection{Average Criterion}
 We assume that the system is stationary, which means that its statistical properties are invariant with respect to time. Therefore, the decision policy $\pi$ for the SMDP model can be defined as a time-invariant mapping from the state space to the action space: $\mathcal{S} \rightarrow \mathcal{A}_s$. Starting from state $s_0$ and continuing with policy $\pi$, the long-term expected average cost can be formulated as follows:
\begin{equation}\label{}
  g^{\pi}(s_0)=\lim_{N\rightarrow\infty}\frac{E_{s_0}^{\pi}\{\sum_{i=0}^N c(s_i,\pi_{s_i})\}}{E_{s_0}^{\pi}\{\sum_{i=0}^N\tau_i\}},
\end{equation}
where $\tau_i$ is the time interval between the $i$th and $(i+1)$th decision epoch. In our SMDP model, the embedded Markov chain for every available policy is a unichain which consists of a single recurrent class plus a (possible empty) set of transient states. Thus, the average cost is independent on the initial state, namely $g^{\pi}=g^{\pi}(s_0)$, for all $s_0\in\mathcal{S}$. The objective is to find an optimal link selection policy so as to minimize the long-term expected energy cost, i.e.:
\begin{equation}\label{}
  \pi^* \in \arg \min_{\pi} g^{\pi}.
\end{equation}

The Bellman optimality equation is a necessary condition for optimality associated with dynamic programming (DP). The following theorem presents the Bellman optimality equation for a unichain SMDP with average criterion.
\begin{theorem}
For a unichain SMDP, there are a scale $g^*$ and a function of states $v(s)$, $\forall s\in \mathcal{S}$ satisfying the following Bellman optimality equation:
\begin{equation}\label{Bellman}
\begin{split}
  &v(s)=\min_{a \in \mathcal{A}_s}\left\{
                                   c(s,a)-g^*y(s,a)+\sum_{s'\in \mathcal{S}}P_a(s'|s)v(s')
                                 \right\}, \\
  &~~~~~~~~~~~~~~~~~~~~~~~~~~~~~~~~~~~~~~~~~~~~~~~~~~~~~~~~~~~~~\forall~ s \in \mathcal{S}.
\end{split}
\end{equation}
where $g^*$ is the average cost under an optimal policy, $y(s,a)$ is the expected time interval between adjacent decision epochs when action $a$ is taken under state $s$, i.e., $y(s,a)=1/\gamma(s,a)$.
\end{theorem}

A proof of the above theorem can be found in \cite{puterman1994markov}. We have obtained the state transition probability $P_a(s'|s)$ and the expected time interval between adjacent decision epochs $y(s,a)$. Accordingly, a DP-based algorithm (such as the value iteration algorithm or the policy iteration algorithm) derived from the Bellman optimality equation can be used to find an (asymptotically) optimal packet scheduling policy. In general, the discrete-time MDP requires that $y(s,a)$ is identical for each available state-action pair, so that we cannot directly use these algorithms to solve the semi-Markov decision problem with average cost criterion because the expected time interval between adjacent decision epochs $y(s,a)$ is not identical for every state-action pair $(s,a)$.

In order to apply the algorithms of the discrete-time MDP to the SMDP, we should uniformize the event occurrence rates for all state-action pairs by adding extra fictitious decisions. The uniformization method is described as follows:

For all $s\in \mathcal{S}$ and $a\in \mathcal{A}_s$, the uniform constant event occurrence rate $\varphi$ has to satisfy the following inequality:
\begin{equation}\label{}
  [1-P_a(s|s)]\gamma(s,a)\leq \varphi.
\end{equation}
We set $\varphi$ as follows:
\begin{equation}\label{}
  \varphi=\sum_{n=1}^N \lambda_n+\max_r\beta_r,
\end{equation}
and multiply both sides of (\ref{Bellman}) by $\gamma(s,a)/\varphi$. Let
\begin{align*}
&\tilde{c}(s,a)=c(s,a)\gamma(s,a)/\varphi, \\
&\tilde{g}=g^*y(s,a)\gamma(s,a)/\varphi=g^*/\varphi, \\
&\tilde{v}(s)=v(s),
\end{align*}
and
\begin{align*}
  \begin{split}
         \tilde{P}_a(s'|s)=\begin{cases}
                        1-\frac{[1-P_a(s'|s)]\gamma(s,a)}{\varphi},~~s' = s \\
                        \frac{P_a(s'|s)\gamma(s,a)}{\varphi}~~~s' \neq s.
                      \end{cases}
  \end{split}
\end{align*}

Thus, the uniform Bellman optimality equation can be given as:
\begin{equation}\label{}
  \tilde{v}(s)=\min_{a \in \mathcal{A}_s}\left\{
                                   \tilde{c}(s,a)-\tilde{g}+\sum_{s'\in \mathcal{S}}\tilde{P}_a(s'|s)\tilde{v}(s')
                                 \right\},~~\forall~ s \in \mathcal{S}.
\end{equation}

We use the relative value iteration algorithm to obtain the $\epsilon-$optimal policy as shown in Algorithm \ref{alg:RVL}, where the operation symbol $sp$ in step 3 is the span which is defined as $sp(\v)=\max_{s\in \mathcal{S}}v(s)-\min_{s\in \mathcal{S}}v(s)$. If constant $\epsilon$ is small enough, the $\epsilon-$optimal policy can converge to an optimal policy.

\begin{algorithm}[htb]
\caption{Relative Value Iteration Algorithm}
\label{alg:RVL}
\begin{algorithmic} [1]
\STATE Initialize $v^0(s)=0$, for all $s\in\mathcal{S}$, choose fixed index $k^*\in \mathcal{S}$, specify a small constant $\epsilon>0$, and set $m=0$;
\STATE For all $s\in \mathcal{S}$, calculate $v^{m+1}(s)$ as follows:
\begin{small}
\begin{equation}\label{}
  v^{m+1}(s)=\min_{a\in A_s}\left\{
\tilde{c}(s,a)-v^m(k^*)+\sum_{s'\in S}\tilde{P}_a(s'|s)v^m(s')
\right\}.
\end{equation}
\end{small}
\label{step2}
\STATE If $sp({\v}^{m+1}-{\v}^m)<\epsilon$, go to step \ref{step4}; otherwise set $m=m+1$ and go to step \ref{step2}.
\STATE For all $s\in \mathcal{S}$, choose an $\epsilon-$optimal policy as follows:
\begin{small}
\begin{equation}\label{}
  d_{\epsilon}(s)\in\arg\min_{a\in A_s}\left\{
\tilde{c}(s,a)-v^m(k^*)+\sum_{s'\in S}\tilde{P}_a(s'|s)v^m(s')
\right\}.
\end{equation}
\end{small}
\label{step4}
\end{algorithmic}
\end{algorithm}

\subsection{Discounted Criterion}
In practice, the statistical characteristics of the downlink data traffic and the solar radiation are time-variant. For convenience, we assume that they are time-invariant within a finite horizon, e.g., one hour. The discounted SMDP may better formulate the downlink packet scheduling problem, because the decisions in the future will have less impact on the present over time. We assume that the continuous-time discounting rate is $\alpha$ ($\alpha>0$) which means that the present value of one unit received $t$ time units in the future equals $e^{-\alpha t}$. Assuming that the initial state is $s_0$ and the time-invariant policy $\pi$ is followed by, the discounted expected total energy cost can be formulated as:
\begin{equation}\label{}
  v^{\pi}(s_0)=E_{s_0}^{\pi}\left\{\sum_{i=0}^{\infty}e^{-\alpha \sigma_i}c(s_i,\pi_{s_i})\right\},
\end{equation}
where $\sigma_i$ represents the time of the $i$th decision epoch.

We define $Q_a(t,s'|s)$ as the transition probability that the next decision epoch occurs at or before time $t$, and the system state at that decision epoch equals $s'$ under the current state-action pair $(s,a)$. We also define $m_a(s'|s)$ as
\begin{equation}\label{}
  m_a(s'|s)=\int_{t=0}^{\infty}e^{-\alpha t}dQ_a(t,s'|s),
\end{equation}
and $q_a(t,s'|s)$ as the differential coefficient of $Q_a(t,s'|s)$. If the next battery state is $m'=m+k$, the next event occurrence time is $t\in [kT_r, (k+1)T_r)$. Assuming the next event is $e_{n'}$, $q_a(t,s'|s)$ is formulated as follows:
\begin{equation}\label{}
  \begin{split}
    q_a(t,s'|s)=\begin{cases}
                   f_X(t)\int_t^{\infty}f_Y(y)dy, ~t\in [kT_r, (k+1)T_r)\\
                   0,~~\mathrm{others}.
                \end{cases}
  \end{split}
\end{equation}
In the same way, $q_a(t,s'|s)$ for all available $s'$ can be obtained.
Therefore, the discounted expected total energy cost can be also formulated as:
\begin{small}
\begin{equation}\label{}
    \begin{split}
        &v^{\pi}(s_0) \\
        &= c(s_0,\pi_{s_0})+E_{s_0}^{\pi}\left\{e^{\alpha\sigma_1}v^{\pi}(s_1)\right\} \\
        &= c(s_0,\pi_{s_0})+\sum_{s_1\in \mathcal{S}}\left[\int_{\sigma_1=0}^{\infty}e^{-\alpha \sigma_1}dQ_{\pi_{s_0}}(\sigma_1,s_1|s_0)\right]v^{\pi}(s_1) \\
        &= c(s_0,\pi_{s_0})+\sum_{s_1\in \mathcal{S}}\left[\int_{\sigma_1=0}^{\infty}e^{-\alpha \sigma_1}q_{\pi_{s_0}}(\sigma_1,s_1|s_0)d\sigma_1\right]v^{\pi}(s_1) \\
        &= c(s_0,\pi_{s_0})+\sum_{s_1\in\mathcal{S}}m_{\pi_{s_0}}(s_1|s_0)v^{\pi}(s_1).
    \end{split}
\end{equation}
\end{small}
An asymptotically optimal downlink packet scheduling policy satisfies the following Bellman optimality equation:
\begin{equation}\label{}
  v(s)=\min_{a\in \mathcal{A}_s}\left\{c(s,a)+\sum_{s'\in\mathcal{S}}m_a(s'|s)v(s')\right\}.
\end{equation}
According to the Bellman equation, we use the value iteration algorithm (Algorithm \ref{alg:VL}) to solve the discounted semi-Markov decision problem so as to obtain an asymptotically optimal packet scheduling policy.
\begin{algorithm}[htb]
\caption{Value Iteration Algorithm}
\label{alg:VL}
\begin{algorithmic} [1]
\STATE Initialize $v^0(s)=0$, for all $s\in\mathcal{S}$, specify a small constant $\epsilon>0$, and set $m=0$;
\STATE For all $s\in \mathcal{S}$, calculate $v^{m+1}(s)$ as follows:
\begin{equation}\label{}
  v^{m+1}(s)=\min_{a\in \mathcal{A}_s}\left\{c(s,a)+\sum_{s'\in\mathcal{S}}m_a(s'|s)v^m(s')\right\}.
\end{equation}
\label{VIstep2}
\STATE If $sp({\v}^{m+1}-{\v}^m)<\epsilon$, go to step \ref{VIstep4}; otherwise set $m=m+1$ and go to step \ref{VIstep2}.
\STATE For all $s\in \mathcal{S}$, choose an $\epsilon-$optimal policy as follows:
\begin{equation}\label{}
  d_{\epsilon}(s)\in\arg\min_{a\in \mathcal{A}_s}\left\{c(s,a)+\sum_{s'\in\mathcal{S}}m_a(s'|s)v^m(s')\right\}.
\end{equation}
\label{VIstep4}
\end{algorithmic}
\end{algorithm}

\section{Simulation Results}
In this section, we evaluate the system performance of our proposed SMDP-based downlink packet scheduling scheme by Matlab numerical simulation. Specifically, we present the average cost and policy of the relative value iteration algorithm under average criterion, the value iteration algorithm under discounted criterion and the greedy algorithm, respectively. We describe the simplified solar radiation in two states, i.e., direct sunlight and cloud cover. The solar radiation states are sequential and circular. We assume there are two classes of downlink packets. The simulation parameters are summarized in Table \ref{simulation_parameters}. We use Monte Carlo method to generate random data based on the corresponding parameters to measure the average cost. The simulation results of the average cost are averaged over 10 runs, where each simulation run lasts for 3600s.

\begin{table}[H]
\centering
\caption{Simulation Parameters}\label{simulation_parameters}
    \begin{tabular}{|c|c|c|c|}
      \hline
      \textbf{Parameter} & \textbf{value} & \textbf{Parameter} & \textbf{Value} \\
      \hline
      $G_0$ & 50 & $E_{min}$ & 0.05 \\
      \hline
      $G_1$ & 200 & $E$ & 1 \\
      \hline
      $d_0$ & 50 & $\lambda_1$ & 10 \\
      \hline
      $d_1$ & 100 & $\lambda_2$ & 5 \\
      \hline
      $v_w$ & 2 & $\zeta_1$ & 8 \\
      \hline
      $\eta$ & 0.2 & $\zeta_2$ & 10 \\
      \hline
      $\Omega_s$ & 0.1 & $\xi_1$ & 3 \\
      \hline
       $\omega_m$ & 2 & $\xi_2$ & 6 \\
      \hline
       $\omega_s$ & 1.5 & $\epsilon$ & $10^{-10}$ \\
      \hline
      $\alpha$ & $0.05$ & & \\
      \hline
    \end{tabular}
\end{table}

\begin{figure}[H]
  \centering
  \includegraphics[width=7.5cm,height=6cm]{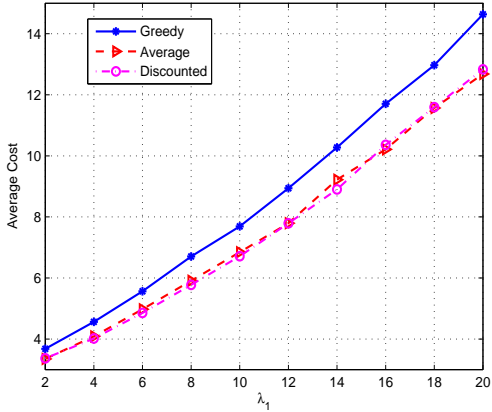}\\
  \caption{The average cost versus the arrival rate of the class $1$ packets}\label{average_cost}
\end{figure}

Fig. \ref{average_cost} presents the average cost versus the arrival rate of the class $1$ packets. The average cost of the relative value iteration algorithm under the average criterion is similar to that of the value iteration algorithm under the discounted criterion. Both of them are less than that of greedy algorithm. In the future work, we will further investigate the performance of the relative value iteration algorithm under the average criterion and the value iteration algorithm under the discounted criterion in slow time-varying systems.

Table \ref{policy} shows the downlink packet scheduling policies for the three algorithms under the default parameters. For a decision-making state $<[r,m],e_n>$, the three columns policies correspond to the relative value iteration algorithm, the value iteration algorithm and the greedy algorithm, respectively. In the greedy algorithm, when a packet arrives, it is served by the SBS as long as there exists enough energy in the battery because of low immediate cost. In the SMDP-based packet scheduling algorithms, the actions are chosen based on the current state and the statistical characteristics of the system.

\begin{table}[H]
\centering
\caption{Policies}\label{policy}
\begin{tabular}{|c||c|c|c||c|c|c||c|c|c||c|c|c|}
    \hline
     m & \multicolumn{3}{|c||}{$<$[0,m],$e_1$$>$} & \multicolumn{3}{|c||}{$<$[1,m],$e_1$$>$} & \multicolumn{3}{|c||}{$<$[0,m],$e_2$$>$} & \multicolumn{3}{|c|}{$<$[1,m],$e_2$$>$}\\
    \hline
    0  & 0 & 0 & 0 & 0 & 0 & 0 & 0 & 0 & 0 & 0 & 0 & 0 \\
    \hline
    1  & 0 & 0 & 0 & 0 & 0 & 0 & 0 & 0 & 0 & 0 & 0 & 0 \\
    \hline
    2  & 0 & 0 & 0 & 0 & 0 & 0 & 0 & 0 & 0 & 0 & 0 & 0 \\
    \hline
    3  & 0 & 1 & 1 & 1 & 1 & 1 & 0 & 0 & 0 & 0 & 0 & 0 \\
    \hline
    4  & 0 & 1 & 1 & 1 & 1 & 1 & 0 & 0 & 0 & 0 & 0 & 0 \\
    \hline
    5  & 0 & 1 & 1 & 1 & 1 & 1 & 0 & 0 & 0 & 0 & 0 & 0 \\
    \hline
    6  & 1 & 1 & 1 & 1 & 1 & 1 & 0 & 0 & 1 & 0 & 0 & 1 \\
    \hline
    7  & 1 & 1 & 1 & 1 & 1 & 1 & 0 & 0 & 1 & 0 & 0 & 1 \\
    \hline
    8  & 1 & 1 & 1 & 1 & 1 & 1 & 0 & 0 & 1 & 0 & 0 & 1 \\
    \hline
    9  & 1 & 1 & 1 & 1 & 1 & 1 & 0 & 0 & 1 & 0 & 1 & 1 \\
    \hline
    10 & 1 & 1 & 1 & 1 & 1 & 1 & 0 & 0 & 1 & 0 & 1 & 1 \\
    \hline
    11 & 1 & 1 & 1 & 1 & 1 & 1 & 0 & 0 & 1 & 0 & 1 & 1 \\
    \hline
    12 & 1 & 1 & 1 & 1 & 1 & 1 & 0 & 0 & 1 & 1 & 1 & 1 \\
    \hline
    13 & 1 & 1 & 1 & 1 & 1 & 1 & 0 & 0 & 1 & 1 & 1 & 1 \\
    \hline
    14 & 1 & 1 & 1 & 1 & 1 & 1 & 1 & 0 & 1 & 1 & 1 & 1 \\
    \hline
    15 & 1 & 1 & 1 & 1 & 1 & 1 & 1 & 0 & 1 & 1 & 1 & 1 \\
    \hline
    16 & 1 & 1 & 1 & 1 & 1 & 1 & 1 & 0 & 1 & 1 & 1 & 1 \\
    \hline
    17 & 1 & 1 & 1 & 1 & 1 & 1 & 1 & 1 & 1 & 1 & 1 & 1 \\
    \hline
    18 & 1 & 1 & 1 & 1 & 1 & 1 & 1 & 1 & 1 & 1 & 1 & 1 \\
    \hline
    19 & 1 & 1 & 1 & 1 & 1 & 1 & 1 & 1 & 1 & 1 & 1 & 1 \\
    \hline
    20 & 1 & 1 & 1 & 1 & 1 & 1 & 1 & 1 & 1 & 1 & 1 & 1 \\
    \hline
\end{tabular}
\end{table}

\section{Conclusions}
In this paper, we proposed an SMDP-based downlink packet scheduling scheme for solar energy assisted HetNets, where the intensity of solar energy is modeled as a CTMC and the arrivals of multi-class downlink packets are modeled as Poisson processes with different rates. We obtained the asymptotically optimal packet scheduling policies with respect to average cost SMDP and discounted cost SMDP. Both the intuitive example and the simulation results show that the asymptotically optimal packet scheduling policies are better than the greedy policy. In our future work, we will consider bandwidth constraints and jointly design the bandwidth allocation and energy management in solar assisted energy HetNets.

\end{document}